# Nonlinear Surface Resistance of $YBa_2Cu_3O_{7-\delta}$ Superconducting Thin Films on MgO Substrates in Dielectric Resonator at Ultra High Frequencies

Dimitri O. Ledenyov

*Electrical and Computer Engineering Department, School of Engineering, James Cook University, Townsville, QLD 4811, Australia*

The nonlinear surface resistance $R_s$ of the $YBa_2Cu_3O_{7-\delta}$ superconducting thin films on the *MgO* substrates was researched in the *Hakki-Coleman* dielectric resonator at the microwave signal powers from *–18 dBm* to *+30 dBm* at the ultra high frequency of *25 GHz* in the range of temperatures from *12 K* to *85 K*. The dependences of the surface resistance on the temperature $R_S(T)$ in the $YBa_2Cu_3O_{7-\delta}$ superconducting thin films at the microwaves were measured. The dependence of the surface resistance on the microwave power $R_S(P)$ in the $YBa_2Cu_3O_{7-\delta}$ superconducting thin films at the microwaves were found at the two temperatures *T = 25 K* and *T = 50 K*. The full expression for the estimation of the measurements accuracy of the surface resistance $R_S$ was derived, and the measurements accuracy was set at *1 %*. The physical mechanisms, which can be used to explain the experimental results, were discussed. It is shown that the surface resistance $R_s$ can nonlinearly increase as a result of the transition by the sub-surface layer of the *HTS* thin film in a mixed state with the *Abricosov* and *Josephson* magnetic vortices generation at an increase of the microwave signal power *P* above the magnitude of *8 dBm*. It is assumed that some additional energy losses have place, because of the microwave power dissipation on the normal metal cores of the magnetic vortices.

## Introduction

The method of the accurate characterization of the superconducting thin films as well as the superconducting single- and poly-crystals in the microwave resonators attracts a considerable research interest, because of a possibility to use this advanced measurement technique during the design optimization of the microwave devices with the superconductors in the microwave circuits in the micro- and nano-electronics. The general approach toward the accurate microwave characterisation of superconductors can be realized, using both the experimental measurements results on the accurate characterization of the high temperature superconductor (*HTS*) thin films at the microwaves in combination with the theoretical formulas to describe the nonlinearities in the high temperature superconductors at the microwaves [1].

Thus, it is necessary to research the dependence of the surface impedance of superconductors $Zs = Rs + jXs$ on the temperature $Zs(T)$, microwave power $Zs\ (P)$ and some other parameters. The present research is aimed to understand the physical phenomena, originating the nonlinearities, and define the nonlinear properties of the $YBa_2Cu_3O_{7-\delta}$ superconducting thin films. Therefore, it would be possible to model the nonlinearities more accurately and apply the computer modeling results during the design of the new advanced electronic devices with the superconductors.

It has to be mentioned that a number of the physical mechanisms, which can be connected with the appearance of the nonlinearities in the *HTS* at the microwaves, were already researched, including the overheating effects, *Josephson phenomena* in the weak links at the inter-grain boundaries, transport of magnetic vortices, de-pairing of the *Cooper* pairs at an action by the current, and some others. However, the universal understanding of the nature of nonlinearities in the superconductors at the microwaves is not achieved yet [2–4].

## Experimental measurements technique

The $YBa_2Cu_3O_{7-\delta}$ superconducting thin films of $10x10\ mm^2$ on the $MgO$ substrates were used to conduct the research on the nonlinear properties of the high temperature superconductors in a *Hakki-Coleman* dielectric resonator with the sapphire rod at the high microwave power levels up to *30 dBm* at the ultra high frequency of *25 GHz* in the range of temperatures from *12 K* to *85 K* as shown in Fig. 1.

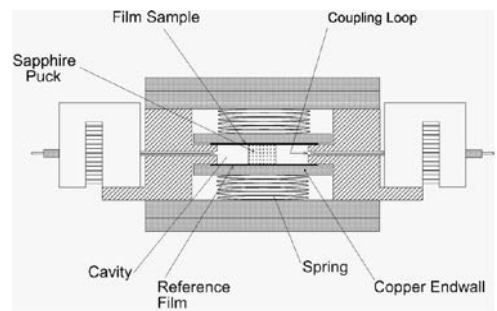

*Fig. 1. Hakki-Coleman dielectric resonator with HTS thin films. Reference and sample thin films are superconductors (after [5]).*

The high quality $YBa_2Cu_3O_{7-\delta}$ superconducting thin films were synthesized by the *THEVA GmbH* in *Germany*. The $YBa_2Cu_3O_{7-\delta}$ superconductor layer thickness is *700 nm* and the $MgO$ substrate thickness is *0.5 mm*. The critical temperature of the *HTS* thin films as stated by the manufacturer is around $T_C = 87K$ and the critical current density at low temperatures is $J_c \sim 2.3 MA/cm^2$.

The ultra high frequency experimental measurement setup to research the nonlinear resonance response of the $YBa_2Cu_3O_{7-\delta}$ thin films on the $MgO$ substrates in a *Hakki-Coleman* dielectric resonator at the microwaves, included the following equipment:
- Vector network analyzer (*HP 8722C*),
- Temperature controller (*Conductus LTC-10*), fitted with a resistive heating element and the two silicon temperature diode sensors,
- Vacuum dewar,
- Close cycle cryogenic laboratory system (*APC-HC4*) suitable for the measurements in a wide range of the temperatures (*10 K– 300 K*),
- Computer system (*IBM-PC*), fitted with a *GPIB* card for the automated control of the temperature controller and network analyzer, as well as the *S*-parameters measurement data transfer from the network analyzer to the computer.

The measurement set up was precisely calibrated by measuring the voltage standing wave ratio *VSWR = (1+|Γ|)/(1−|Γ|)*, (*Γ* is the reflection coefficient), before the accurate experimental measurements of the *S*-parameters as it is schematically shown in Fig. 2.

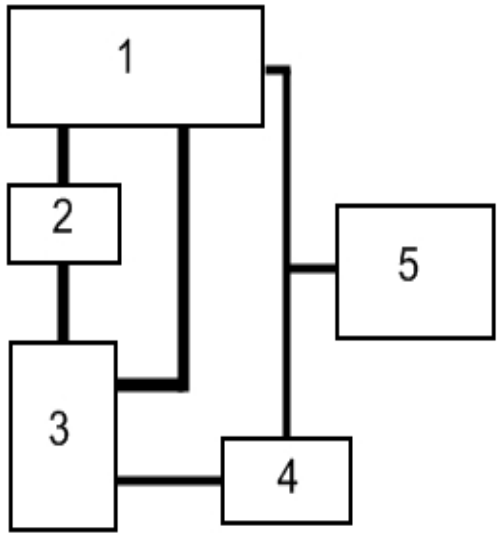

*Fig. 2. Block scheme of the experimental for surface resistance measurements of HTS films: 1− vector analyzer HP 8722C, 2− RF amplifier, 3− cryogenic system with dielectric resonator, 4− temperature controller, 5− computer system.*

The *Hakki-Coleman* dielectric resonator (*HCDR*) and the microstrip resonator for the accurate characterization of the *HTS* thin films at microwaves were used. The loaded $Q_L$-factor and the coupling coefficients $\beta_1$ and $\beta_2$ of a dielectric resonator were obtained from the multi-frequency measurements of the $S_{21}$, $S_{11}$ and $S_{22}$ parameters, which were measured around the resonance, using the *Transmission Mode Q-Factor* (*TMQF*) technique [6, 7]. The *TMQF* method enables us to obtain the accurate values of the surface resistance, accounting for the factors such as the noise, delays due to the uncompensated transmission lines, and cross-talks, occurring in the measured data. The unloaded $Q_0$-factor was calculated from the exact equation (1)

$$Q_0 = Q_L(1+\beta_1+\beta_2), \qquad (1)$$

using the *TMQF* method at all the temperatures.

The *Hakki-Coleman* dielectric resonator with the embedded *HTS* samples was mounted inside the vacuum dewar and cooled down to the temperature *T* of around *12 K*, and the *S*-parameters were measured at the resonance frequency up to the temperature *T = 85 K*. The microwave power of the source signal was *–5 dBm* and the number of points was *1601*.

The surface resistance $R_S$ of $YBa_2Cu_3O_{7-\delta}$ superconducting thin films on the $MgO$ substrates has been computed, using the software *SUPER* [8], based on the equation (1), as shown in the expression in eq. (2)

$$R_S = A_S \left\{ \frac{1}{Q_0} - \frac{R_m}{A_m} - p_e \tan\delta \right\}. \qquad (2)$$

The geometric factors $A_S$, $A_m$, and $p_e$ were computed, using the incremental frequency rules as written in eq. (3) [9]

$$A_S = \frac{\omega^2 \mu_0}{4} \bigg/ \frac{\partial \omega}{\partial L}, \quad A_m = \frac{\omega^2 \mu_0}{2} \bigg/ \frac{\partial \omega}{\partial a},$$
$$p_e = 2\left|\frac{\partial \omega}{\partial \varepsilon}\right|\frac{\varepsilon_r}{\omega} \qquad (3)$$

The formulas allow us to get magnitudes of surface resistance $R_S$ of the $YBa_2Cu_3O_{7-\delta}$ superconducting thin films at the microwaves, and research the dependences of the surface resistance on the temperature $R_S(T)$, and the dependences of the surface resistance on the microwave power $R_S(P_{rf})$. In the case of the dielectrics, it is possible to research the characteristics of the dielectrics, including the influence by the microwave power *P* on the dielectrics.

### Dependence of surface resistance on temperature $R_S(T)$ in $YBa_2Cu_3O_{7-\delta}$ superconducting thin films at microwaves

The six samples with the $YBa_2Cu_3O_{7-\delta}$ superconducting thin films on the $MgO$ substrates have been used during the accurate characterization of their physical properties at the microwaves. All the $YBa_2Cu_3O_{7-\delta}$ superconductor thin films were divided into the two groups and the measurements were performed in the pairs on the "*round robin*" rotation basis to enable the determination of the microwave parameters of an every *HTS* sample. For the first group of the three samples, the temperature dependences of the sums of the surface resistances $R_{S1}+R_{S2}$, $R_{S1}+R_{S3}$, $R_{S2}+R_{S3}$ were obtained in the process of measurements in the temperature range from *12 K* to *85 K*. The dependences of the surface resistance on the temperature: $R_{S1}(T)$, $R_{S2}(T)$ and $R_{S3}(T)$ were derived from the above data. The same measurements were conducted with the second group of samples. The obtained data for the dependences of the surface resistances on the temperature for an every tested sample are shown in Figs. 3 and 4.

Analyzing the measurements results, it can be seen that the values of the residual surface resistance $R_S$ in the $YBa_2Cu_3O_{7-\delta}$ superconductor thin films are different for the every measured sample. The changes of the residual surface resistance in the $YBa_2Cu_3O_{7-\delta}$ superconducting thin films at the temperature of *12 K* can be connected with the quality of microwave samples [6]. The *sample 1* has the lowest surface resistance $R_S$ in the entire temperature range with the residual surface resistance $R_{Sr}$ values in the range of *0.3–0.5 mΩ*, while the sample 2 and sample 3 have much higher values of up to *1.8–2 mΩ* at the temperature of *75 K*. The *sample 6* in the second bunch of the $YBa_2Cu_3O_{7-\delta}$ superconducting thin films has the values of the surface resistance $R_{Sr}$ with the lowest surface resistance $R_{Sr}$ for this group in Fig. 4, but it is still higher than in the case of the *sample 1* in Fig. 3.

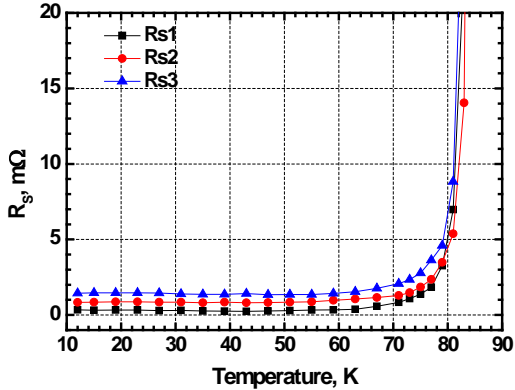

***Fig. 3.*** *Dependence of surface resistance on temperature $R_s(T)$ in $YBa_2Cu_3O_{7-\delta}$ thin films on MgO substrate (samples 1-3) at microwave signal power of -5dBm at frequency of 25GHz.*

The dependence of the surface resistance on the temperature $R_S(T)$ may be represented at the temperatures below the critical temperature $T < T_c$ as in eq. (4)

$$R_S(T) \approx R_{Sr} + R_n(T_C)\exp(-\Delta(T)/k_BT), \qquad (4)$$

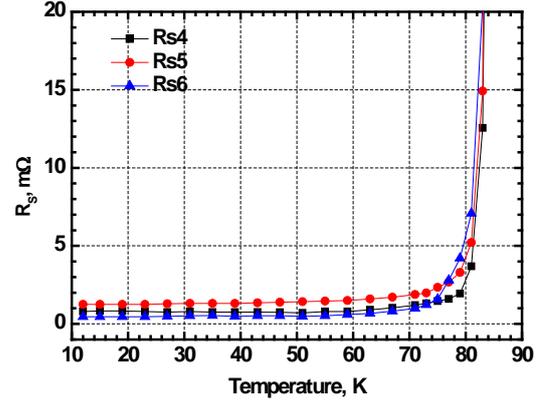

***Fig. 4.*** *Dependence of surface resistance on temperature $R_s(T)$ in $YBa_2Cu_3O_{7-\delta}$ thin films on MgO substrate (samples 4–6) at microwave signal power of -5dBm at frequency of 25GHz.*

where $R_{Sr}$ is the residual resistance at the low temperatures $T$, $R_n$ is the surface resistance in a normal state at $T \geq Tc$, $\Delta(T)$ is the temperature dependent *BCS* superconducting energy gap, $k_B$ is the *Boltzmann* constant. The typical experimental *(1)* and modeled *(2)* dependences of the surface resistance on the temperature $R_S(T)$ are shown in Fig. 5.

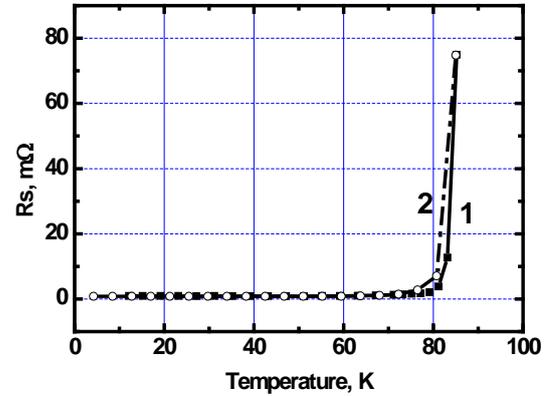

***Fig. 5.*** *Experimental (1, square) and modeled (2, circle) dependences of surface resistance on temperature $R_S(T)$ for measured surface resistance $R_{S1}$ in $YBa_2Cu_3O_{7-\delta}$ thin film in Fig. 3.*

In Fig. 5, the *curve 2* was modeled, using the eq. (4) with $\Delta(T) \approx 3\Delta_{BCS}(T)$, $R_{Sr} \approx 0.7\ m\Omega$ and $R_n \approx 78\ m\Omega$. Author used the standard *Barden, Cooper, Schrieffer* (*BCS*) theory dependence of the energy gap on the temperature $\Delta_{BCS}(T)$. In Figs. 3 and 4, the parameters: $\Delta(T)$, $R_{Sr}$ and $R_n$ can be found for the each curve. These parameters: $\Delta(T)$, $R_{Sr}$ and $R_n$ differ slightly, because the $YBa_2Cu_3O_{7-\delta}$ superconducting thin films have the different qualities of synthesis.

## Dependence of surface resistance on microwave power $R_S(P)$ in $YBa_2Cu_3O_{7-\delta}$ superconducting thin films at microwaves

The accurate characterization of the dependences of the surface resistance on the microwave power $R_S(P)$ in the $YBa_2Cu_3O_{7-\delta}$ superconducting thin films on the $MgO$ substrates were conducted, utilizing the measurement system in Fig. 2. The additional microwave power amplifier, connected to the input port of a microwave resonator under the test inside the vacuum dewar, enabled author to conduct the measurements from the $-18\ dBm$ to $+30\ dBm$ microwave signal power levels. It should be mentioned that at the higher microwave power levels, the attenuator was placed at output port of a microwave resonator to prevent the high microwave power signals entering the input port of the vector network analyzer. The attenuator was able to attenuate the microwave signal power on the $-20\ dBm$. The amplifier had the microwave signal amplification range from $+20\ dBm$ to $+25\ dBm$, and enabled us to perform the measurements from $0\ dBm$ to the maximum of $+30\ dBm$. The appropriate scale modifications were made to the output signal measurements results in order to numerically compensate for the attenuation effect.

The experimental measurements results for the dependences of the surface resistance $R_S$ vs. the microwave power $P$ are presented in Figs. 6 and 7.

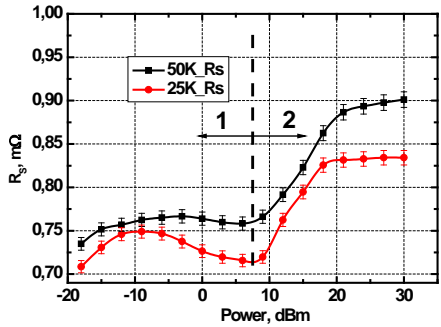

***Fig. 6.** Dependence of surface resistance on microwave power $R_S(P)$ in $YBa_2Cu_3O_{7-\delta}$ thin films on MgO substrate at frequency of 25 GHz at temperatures $T = 25\ K$ and $T = 50\ K$, samples (1-3).*

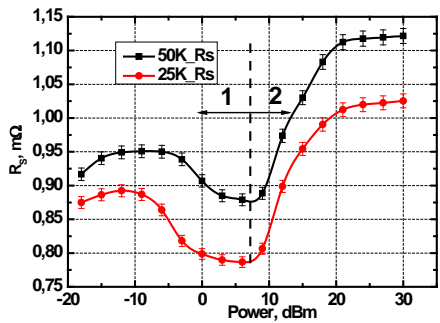

***Fig. 7.** Dependence of surface resistance on microwave power $R_S(P)$ in $YBa_2Cu_3O_{7-\delta}$ thin films on MgO substrate at frequency of 25 GHz at temperatures $T = 25\ K$ and $T = 50\ K$, samples (2-3).*

Fig. 6 shows that the applied microwave power $P$ changes the value of the surface resistance $Rs$ in the $YBa_2Cu_3O_{7-\delta}$ superconducting thin films (*samples 1 - 3*) at the temperatures $T = 25\ K$ and $T = 50\ K$. As it can be seen, the pair of the $YBa_2Cu_3O_{7-\delta}$ superconducting thin films (*samples 1 - 3*) has a nonlinear dependence of the surface resistance on the microwave power $R_S(P)$ at the temperatures of $25\ K$ and $50\ K$. The minimum in the dependence $R_S(P)$ is observed, when the magnitude of applied microwave power is $P \approx 10\ dBm$ approximately.

Fig. 7 shows the nonlinear dependence of the surface resistance on the microwave power $R_S(P)$ in the $YBa_2Cu_3O_{7-\delta}$ superconductor thin films (*samples 2 − 3*). In the range of the microwave signal powers of $0 − 10\ dBm$, the surface resistance $R_S$ has a minimum similar to other samples in agreement with the research data in [10 − 14], considered for the two level systems.

In Fig. 6 and Fig. 7, it can be noticed that there are the two different regions, where the $YBa_2Cu_3O_{7-\delta}$ superconducting thin films samples are exposed to 1) the low microwave power external magnetic field of the ultra high frequency $H_{rf} < H_{C1}$ (*region 1*) and 2) the high microwave power external magnetic field of the ultra high frequency $H_{rf} > H_{C1}$ (*region 2*). The computing of the magnitudes of the surface resistances $R_S$ were conducted in the *region 1* and *region 2* of the $YBa_2Cu_3O_{7-\delta}$ superconducting thin films at the microwaves, using the eq. (2). Author would like to make an emphasis on the different physical mechanisms of the energy losses by the electromagnetic waves, existing in the *region 1* and the *region 2* in the $YBa_2Cu_3O_{7-\delta}$ superconducting thin films at the microwaves. These physical mechanisms will be comprehensively discussed below.

### Discussion on measurements accuracy

The accuracy of measured results is an important issue to consider, when conducting the novel research projects. In order to identify the range of acceptable variations in the obtained data, the theoretical analysis is made in reference to the equations and parameters, related to the obtained results.

The surface resistance has been described as in eq. (5)

$$R_S = A_S \left\{ \frac{1}{Q_0} - \frac{R_m}{A_m} - \frac{1}{Q_d} \right\}, \quad (5)$$

where $Q_d = 1/p_e \tan\delta$.

In order to determine the magnitude of the correction to the $R_S$ value, i.e. the $\Delta R_S$ or relative $\Delta R_S/R_S$, it is necessary to represent all the values in the formula (5) with their corresponding independent $\Delta$ corrections, as shown in eq. (6)

$$R_S + \Delta R_S = (A_S + \Delta A_S)\left( \frac{1}{Q_0 + \Delta Q_0} - \frac{R_m + \Delta R_m}{A_m + \Delta A_m} - \frac{1}{(p_e + \Delta p_e)(\tan\delta + \Delta\tan\delta)} \right) \quad (6)$$

To eliminate the fractions in the denominators, the expression $(A_S + \Delta A_S)$ can be modified as $A_S(1 + |\Delta A_S / A_S|)$ in eq. (7)

$$\frac{1}{Q_0 + \Delta Q_0} = \frac{1}{Q_0 \left(1 + \frac{\Delta Q_0}{Q_0}\right)} = \frac{1}{Q_0}\left(1 - \frac{\Delta Q_0}{Q_0}\right) = \frac{1}{Q_0}\left(1 + \left|\frac{\Delta Q_0}{Q_0}\right|\right) \quad (7)$$

It follows from the rule that $1/(1+\Delta x) = 1 - \Delta x$, which is valid, if $x \ll 1$. In view of the fact that the $\Delta x$ may be $\pm |\Delta x|$, author uses only $+|\Delta x|$ so that the error has a maximum value.

The second part of the equation (6) will look like in eq. (8)

$$\frac{R_m + \Delta R_m}{A_m + \Delta A_m} = (R_m + \Delta R_m)\frac{1}{A_m}\frac{1}{1 + \frac{\Delta A_m}{A_m}} = R_m\left(1 + \left|\frac{\Delta R_m}{R_m}\right|\right)\frac{1}{A_m}\left(1 + \left|\frac{\Delta A_m}{A_m}\right|\right) \quad (8)$$

and the last term can simply be written as in eq. (9)

$$(p_e + \Delta p_e)(\tan \delta + \Delta \tan \delta) = \frac{1}{Q_d}\left(1 + \left|\frac{\Delta p_e}{p_e}\right|\right)\left(1 + \left|\frac{\Delta \tan \delta}{\tan \delta}\right|\right) \quad (9)$$

After the substitutions and multiplications, all the related terms, which are higher than first degree of $\Delta$, can be discarded as in eq. (10)

$$1 + \left|\frac{\Delta R_S}{R_S}\right| = \frac{A_S}{R_S}\left(1 + \left|\frac{\Delta A_S}{A_S}\right|\right)\left[\begin{array}{c}\frac{1}{Q_0}\left(1 + \left|\frac{\Delta Q_0}{Q_0}\right|\right) - \frac{R_m}{A_m}\left(1 + \left|\frac{\Delta R_m}{R_m}\right|\right)\left(1 + \left|\frac{\Delta A_m}{A_m}\right|\right) \\ -\frac{1}{Q_d}\left(1 + \left|\frac{\Delta p_e}{p_e}\right|\right)\left(1 + \left|\frac{\Delta \tan \delta}{\tan \delta}\right|\right)\end{array}\right] \quad (10)$$

It can be observed that the $|\Delta R_S / R_S|$ is a function of the six independent variables $\Delta A_S$, $\Delta Q_0$, $\Delta R_S$, $\Delta A_m$, $\Delta p_e$ and $\Delta \tan \delta$.

Now, the coefficients for these variables need to be found as well as the expression for the term $|\Delta R_S / R_S|$ as a multidimensional vector, where the mentioned independent variables are the vectors, needs to be formulated. To calculate the coefficients, the term $|\Delta R_S / R_S|$ will be present in the left side of the formula (8). In the right side, only a variable of the interest will be left, and all the others will supposed to be the zeros. Thus, a sum of the six terms will be obtained in the right part, where each term depends on its own variable only, as shown in eq. (11).

$$\left|\frac{\Delta R_S}{R_S}\right| = C_1\left|\frac{\Delta A_S}{A_S}\right| + C_2\left|\frac{\Delta Q_0}{Q_0}\right| + C_3\left|\frac{\Delta R_m}{R_m}\right| + C_4\left|\frac{\Delta A_m}{A_m}\right| + C_5\left|\frac{\Delta p_e}{p_e}\right| + C_6\left|\frac{\Delta \tan \delta}{\tan \delta}\right| \quad (11)$$

It follows from eq. (10) that the $C_1$, $C_2$, $C_3$, $C_4$, $C_5$ can be written as in eq. (12)

$$\begin{aligned} C_1 &= \left(\frac{A_S}{R_S}\right)\left(\frac{1}{Q_0} - \frac{R_m}{A_m} - \frac{1}{Q_d}\right), \\ C_2 &= \left(\frac{A_S}{R_S Q_0}\right), \\ C_3 &= C_4 = \left(\frac{R_m A_S}{R_S A_m}\right), \\ C_5 &= C_6 = \left(\frac{A_S}{R_S Q_d}\right). \end{aligned} \quad (12)$$

In the multidimensional space, the vector length square is equal to a sum of the squares of the orthogonal vectors in eq. (13)

$$\left|\frac{\Delta R_S}{\Delta R}\right|^2 = C_1^2\left|\frac{\Delta A_S}{A_S}\right|^2 + C_2^2\left|\frac{\Delta Q_0}{Q_0}\right|^2 + C_3^2\left|\frac{\Delta R_m}{R_m}\right|^2 + C_4^2\left|\frac{\Delta A_m}{A_m}\right|^2 + C_5^2\left|\frac{\Delta p_e}{p_e}\right|^2 + C_6^2\left|\frac{\Delta \tan \delta}{\tan \delta}\right|^2 \quad (13)$$

or as written in eq. (14)

$$\left|\frac{\Delta R_S}{R_S}\right|^2 = \left(\frac{A_S}{R_S}\right)^2\left(\frac{1}{Q_0} - \frac{R_m}{A_m} - \frac{1}{Q_d}\right)^2\left|\frac{\Delta A_S}{A_S}\right|^2 + \left(\frac{A_S}{R_S Q_0}\right)^2\left|\frac{\Delta Q_0}{Q_0}\right|^2 + \left(\frac{R_m A_S}{R_S A_m}\right)^2\left(\left|\frac{\Delta R_m}{R_m}\right|^2 + \left|\frac{\Delta A_m}{A_m}\right|^2\right) + \left(\frac{A_S}{R_S Q_d}\right)^2\left(\left|\frac{\Delta p_e}{p_e}\right|^2 + \left|\frac{\Delta \tan \delta}{\tan \delta}\right|^2\right) \quad (14)$$

The magnitude of $|\Delta R_S / R_S|$ is equal to the square root of the right part of the eq. (13).

The next step requires the making of certain assumptions about the magnitudes of the involved terms. The $\Delta A_S$, $\Delta A_m$ and $\Delta p_e$ are measured with the accuracy $0.01$, that is to say $1\%$; the $\Delta \tan \delta$ has a very small value, because the $\tan \delta$ is very small for the dielectrics $(10^{-6} - 10^{-8})$ at the low temperatures and the $Q_d$ is very large $\sim 10^7$.

Therefore, the $C_5$ and $C_6$ are very small, and these members make a very little contribution to the error of the experiment and they can be disregarded. It is also possible to discount both the $\Delta \tan \delta$ value and all the elements with the $\Delta \tan \delta$. In this case, the first term is determined by the fact that the uncertainty (error) $\Delta A_S$, which is equal to $0.01$, is divided by $Q_0$, which is equal to around $10^4$, that might be even bigger for the dielectric resonators. It is clear that this error can be neglected. The second term can be disregarded as well, because it includes $Q_0$ in the power of the two in the denominator. The following element of the equation will be proportional to the $\Delta A_S$, and consequently, is around $0.01$. The fourth term is proportional to $\Delta A_m$ and is $0.01$ approximately. Therefore, the following two members are present in the $\Delta R_S$ expression in the eq. (15)

$$\left|\frac{\Delta R_S}{R_S}\right| \approx \left[\left(\frac{A_S}{R_S}\right)^2 \left(\frac{R_m}{A_m}\right)^2 \left|\frac{\Delta A_S}{A_S}\right|^2 + \left(\frac{R_m A_S}{R_S A_m}\right)^2 \left(\left|\frac{\Delta R_m}{R_m}\right|^2 + \left|\frac{\Delta A_S}{A_S}\right|^2\right)\right]^{1/2}$$
$$= \left[\left(\frac{R_m A_S}{R_S A_m}\right)^2 \left(\left|\frac{\Delta A_S}{A_S}\right|^2 + \left|\frac{\Delta R_m}{R_m}\right|^2 + \left|\frac{\Delta A_S}{A_S}\right|^2\right)\right]^{1/2} \quad (15)$$

The biggest error will be in the case, when all the elements values are added, having taken them by the modulo. It is clear that the main error and uncertainty in the $Rs$ magnitude depends on the $\Delta A_m$ и $\Delta A_S$ (their values are expressed in the *ohms* – same as for the *R*), i.e. on the geometrical factors of the normal metal and superconductor, which are determined by the geometry of both the sample and the resonator, and do not depend on the signal level at the first approximation.

It is assumed that the geometrical factors do not depend on the microwave power or other parameters in the formula at the certain temperatures. Then, the absolute value of the correction does not depend on the microwave power either, and is equal to the total value of the absolute corrections.

From the analysis, it is evident that the error values will not have practical grounds to exceed a certain range of around *1 %* in the researched case. The error band limits at *1 %* are set in Figs. 6 - 7.

### Discussion on measurements results

In the *HTS* films, there may be several reasons for the nonlinear dependence of the surface resistance on the microwave power $Rs(P)$ [1]. It is a well known fact that there are many factors, which may contribute to the nonlinear dependence $Rs(P)$ such as the thermal heating, *Josephson* vortex motion in the inter-grain weak links, nonlinear dynamics of the *Josephson* vortices in the high-$J_c$ *Josephson* contacts, nonlinear resistively shunted junctions, etc. The *Josephson* junction model may be appropriate for the *HTS* thin films with the not good enough inter-grain contacts [15]. These contacts may contribute to the increase of the magnitude of the residual surface resistance $R_{Sr}$ at the low microwave power levels at the low temperatures [16]. In the high quality superconductors, when the *Josephson's* contacts are absent, the appearance of the nonlinearities may be connected with the *Abricosov* vortices, and their dynamics in the *HTS*, when the magnetic field of microwaves reaches the magnitudes of $H_{rf} \geq H_{C1}$, where $H_{C1}$ is the low critical magnetic field of the *HTS*. Author developed a model of the nonlinear surface resistance $R_S$, using the dependence of the surface resistance on the energy gap $Rs(\Delta(T))$, which follows from the *BCS* theory for the traditional superconductors with the *s*-type symmetry of the superconducting electron pairs wave function (see, [1]). In this case, the energy gap value does not depend on the orientation in the momentum space, and it is a function from the temperature only. As author shows in Fig. 5, the *BCS* like eq. (4) may be used to explain the temperature dependence of the surface resistance $R_S(T)$.

In the *HTS* superconductors, the wave function of the electron pairs has the *d*-symmetry or the (*s* + *d*)- symmetry [1]. The electromagnetic field penetrates into the surface layer of the superconductor on the penetration depth $\lambda \approx 10^{-5}$–$10^{-6}$ $cm$, and it decreases in the $YBa_2Cu_3O_{7-\delta}$ superconductor thin film by the exponential law. In the *HTS*, the *Ginzburg-Landau* parameter is $\kappa = \lambda/\xi \gg 1/\sqrt{2}$, where $\xi \approx 10^{-7}$ $cm$ ($\xi$ is the coherence length or the dimension of the superconducting pair of electrons with the momentums $p_1$ and $p_2$). These superconductors are of the second type (*London* type), and their electrodynamics properties are described by the local *London* theory.

The full momentum of an electron pair is equal to zero, and $p_1 = -p_2$, where $|p_{1,2}| \sim |p_F|$. Hence, the full velocity of an electron pair is zero too, but every electron has the momentum $p_{1,2} \sim p_F$. In the real space in the magnetic field, the electron moves along the spiral trajectory and rotates around the magnetic field direction in the plane, which is perpendicular to the magnetic field $H$. Every electron in the superconducting electron pair has the correlating electron with the opposite momentum on the distance of around $\xi$. In the coordinate system, which is connected with the centre of the electron pair mass, the momentums of electrons must rotate with the cyclotron frequency $\omega_C = eB/m_e$, which, in the magnetic field $H_{C1} \approx 8000$ $A/m$ (~$100$ $Oe$), is only $\omega_C \approx 1,76 \cdot 10^9$ $rad/sec$ or $\omega_C/2\pi \approx 2,8 \cdot 10^8$ $Hz$, and the cyclotron period $\tau_C \approx 3,6 \cdot 10^{-9}$ $sec$ ($m_e$ is the electron mass). The cyclotron frequency is much smaller than the electromagnetic field frequency of $2,5 \cdot 10^{10}$ $Hz$, and it is not possible to account for the electrons momentum rotation in the momentum space at the action by the magnetic field of the electromagnetic wave. The electron mean free path $l$ in the *HTS* at the low temperature $T$ depends on both the quality of material structure and the purity of the *HTS*, and the $l$ may be of a few *mm* in the high quality crystal with the big life time of electron $\tau_e \approx l/v_F$. At the high temperature, which is close to the critical temperature $T_C$, the electron mean free path $l$ is defined by the mechanism of the electrons scattering on the phonons, and it has a small length of around a few nanometers.

In the both cases, the dependence of the alternate magnetic field on the frequency $H(\omega, t)$ produces the alternate electrical field $rot\, E_t = dH_t/dt$ and $E_t \sim \omega\mu\mu_0 H_t$, which acts on the normal and superconducting electrons, generating the surface electrical current density $J_S \sim \sigma_S E_t$, where $\sigma_S$ is the surface conductivity ($E_t$ and $H_t$ are the tangential components of the electromagnetic field, $\mu \approx 1$). The *Leontovich* boundary conditions $E_t = Z_S [H_t \times n]$ are true in the normal metals and superconductors ($n$ is the normal vector, which is perpendicular to the surface and oriented toward the metal). The complex surface impedance is equal to $Z_S = R_S + jX_S = j\omega\mu_0\lambda_{SC}$, where $X_S$ is the surface reactance and $\lambda_{SC}$ is the complex penetration depth by the magnetic field into the *HTS*.

As shown above, the dependence of the surface resistance on the temperature $R_S(T)$ can be written as in eq. (4). The presence of the residual resistance $R_{Sr}$, which has a big enough magnitude in the researched *HTS* samples, can be connected with the fact that the superconductor has the energy gap $\Delta(T, \varphi)$, which is a function of the angle $\varphi$ in the momentum space in the *ab-plane*, and it has the *d-wave* symmetry in the momentum space, $\Delta_d(\varphi) = \Delta_0 cos2\varphi$, as shown in Fig. 8. In this case, the second term, $R_n \cdot exp(-\Delta(T,\varphi)/kT)$, in eq. (4) must be averaged on the $\varphi$ at the given temperature *T*. The shapes of the dependence of the surface resistance on the angle $R_S(\varphi)$ at the two temperatures *T = 85K (1), T = 25K (2)* are shown in Fig. 9.

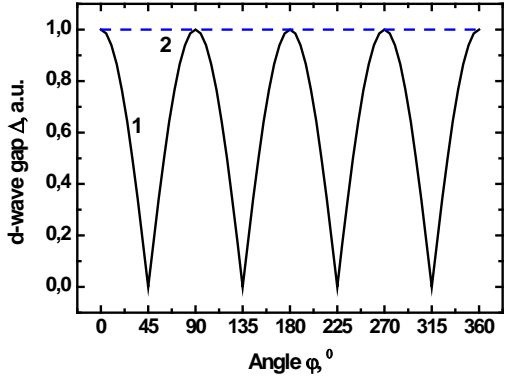

*Fig. 8. Dependence of superconducting d-wave energy gap on angle $\Delta_d(\varphi) = \Delta_0 cos2\varphi$ (1) and constant s-type energy gap $\Delta$ (2) in momentum space in ab-plane.*

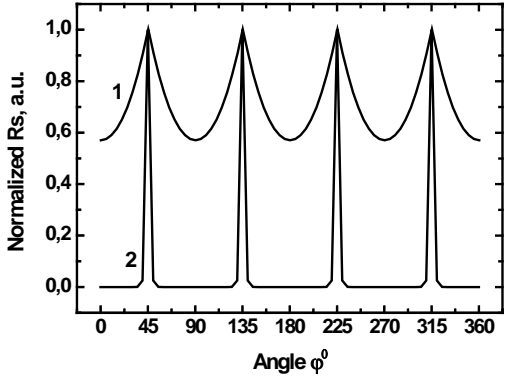

*Fig. 9. Dependence of normalized d-wave surface resistance on angle $R_S(\varphi)$ in momentum space in ab-plane at temperatures T=85K (1) and T=25K (2).*

Thus, it is possible to perform the accurate physical characterization of the dependence of the surface resistance on the temperature $R_S(T)$ in the $YBa_2Cu_3O_{7-\delta}$ superconducting thin films on the various substrates in the frames of the developed theoretical representation. Let us consider the experimental measurements results on the dependence of the surface resistance on the microwave power *Rs(P)*. The magnitude of the dissipated microwave power of an electromagnetic wave in an *HTS* sample is $P_S \propto J_S^2 R_S$ and $R_S \propto 1/Q$. In this case, the *HTS* sample can be modeled as a resistor, which is embedded in the serial lumped elements circuit to model the metals and superconductors [1] as it is done in this research. Let us add that the magnitude of the dissipated microwave power of an electromagnetic wave in the *HTS* sample has to be expressed as $P_S \propto E_S^2/R_S$ in the case of the certain parts of an *HTS* sample, which transit to the dielectric state at an action by the applied electromagnetic field, such as the *Josephson* contacts on the inter-grain boundaries in the crystal lattice or the impurities with the non-stochastic phase with the small contents of *oxygen* in the $YBa_2Cu_3O_{7-\delta}$ superconducting thin film. In this case, the magnitude of dissipated energy could be smaller, if the magnitude of resistivity of a resistor would be bigger, and the surface resistance $R_S$ is proportional to the quality factor $R_S \propto Q$. This conclusion results in a necessity to model this part of a *HTS* sample with the application of the parallel equivalent lumped elements resonance circuit. In the researched samples, it is necessary to consider that the energy of the electromagnetic wave mainly dissipates on the *Josephson* contacts, which are created at the inter-grain boundaries in the crystal lattice of the $YBa_2Cu_3O_{7-\delta}$ superconducting thin film. At the same time, the *Abricosov* magnetic vortices, which originate in the $YBa_2Cu_3O_{7-\delta}$ superconducting thin film at $H_{rf} \geq H_{C1}$, make a main contribution to the energy dissipation in the *HTS* thin film in the range of the high external magnetic fields [1]. The transition region from the one theoretical mechanism of energy dissipation to another theoretical mechanism of energy dissipation is denoted as a vertical dashed line in Figs. 6 and 7. Therefore, it is possible to assume that the effective magnitude of the surface resistance $R_S^*$ in the $YBa_2Cu_3O_{7-\delta}$ superconducting thin film in the region *1* at the low magnetic fields must be opposite to the given value, that is $R_S^* \propto 1/R_S$ in relation to the point of *8 dBm*, where some other physical mechanism comes to the action as shown in Fig. 10 for the data in Fig. 7.

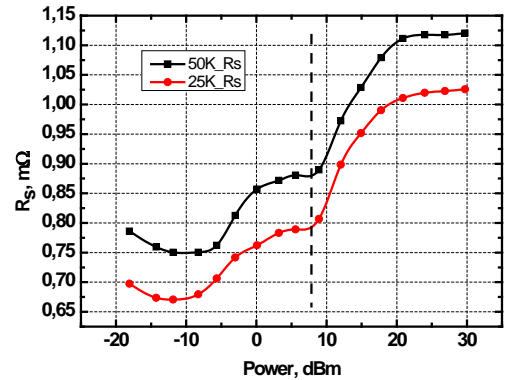

*Fig. 10. Microwave power dependence of surface resistance $R_S(P)$ of $YBa_2Cu_3O_{7-\delta}$ thin films on MgO substrate at frequency of 25 GHz at temperatures T = 25 K, 50 K, samples (2-3), applying modeling with parallel lumped elements equivalent resonance circuit in range of small magnitudes of microwave power.*

Considering the dependences in Fig. 10, it is possible to assume that the *Josephson* magnetic vortices originate in the *Josephson* contacts at the inter-grain boundaries in the crystal lattice of the $YBa_2Cu_3O_{7-\delta}$ superconducting thin films, when the magnitude of the external magnetic field $H_e$ reaches the magnitude of the lower critical magnetic field $H_{C1J}$ at the applied microwave power $P$ level of $-10$ $dBm$, resulting in an increase of the energy dissipation in a *HTS* sample. The considered physical mechanisms acts up to the microwave power level of *+5 dBm*, reaching its maximum. The subsequent increase of the magnitude of the microwave power up to the magnitude of the lower critical magnetic field $H_{C1}$ results in an appearance of the *Abricosov* magnetic vortices in the $YBa_2Cu_3O_{7-\delta}$ superconducting thin film. Author uses the conditional terminology in regards to the *Abricosov* magnetic vortices, because the transition of the surface layer of the $YBa_2Cu_3O_{7-\delta}$ superconducting thin film to a mixed state with the *Abricosov* magnetic vortices generation has its distinctive features, comparing to the case of the transition of a bulk *HTS* sample to a mixed state, when the external magnetic field is higher than the lower critical magnetic field $H_e > H_C$. In the considered case, the external alternate magnetic field $H_e$ only penetrates into the sub-surface layer in the $YBa_2Cu_3O_{7-\delta}$ superconducting thin film on the penetration depth $\lambda$, and the normal metal cores of the *Abricosov* magnetic vortices with the diameter of $2\xi$ begin to originate at $H_{rf} > H_{C1}$. This process and its features are discussed in details below.

As far as the data on the surface resistance $R_S$ in the $YBa_2Cu_3O_{7-\delta}$ superconducting thin films in the range of the high microwave powers of above *8 dBm* is concerned, the high temperature superconductors (*HTS*) are classified as the local superconductors of the *London* type, hence their physical properties can be characterized by the average parameters in the local regions, having the dimensions, which are comparable with the coherent length dimension $\xi$.

Author considers the two possible cases no. 1 and 2:

1. The $YBa_2Cu_3O_{7-\delta}$ superconducting thin film is under the action by the **constant external magnetic field $H_e$**, and the *Abricosov* magnetic vortices penetrate inside the *HTS* sample in the range of the magnitudes of the external magnetic fields $H_e \geq H_{C1}$. In this case, the external magnetic field $H_e$ is constant, and there is no the transport current in the *HTS* sample, hence the *Abricosov* magnetic vortices are fixed at their positions, and it looks like the *Abricosov* magnetic vortices must not contribute to the energy dissipation in the $YBa_2Cu_3O_{7-\delta}$ superconducting thin film at microwaves. However, it makes sense to explain that the *Abricosov* magnetic vortices have the normal metal cores with the radiuses, which are equal to the coherent length $\sim \xi$ approximately, and the penetrating constant external magnetic field $H_e$ excites the alternating electromagnetic fields and currents in the layer with the thickness of $\lambda$ in the $YBa_2Cu_3O_{7-\delta}$ superconducting thin film, resulting in the additional energy dissipation on the normal electron excitations in the normal metal cores of the magnetic vortices. In the range of the high constant magnetic fields up to $B_e \sim 15\ T$, when the total number of the *Abricosov* magnetic vortices can be precisely counted, this physical mechanism of the electromagnetic wave scattering on the electron excitations in the normal metal cores was observed and researched before in [18,19].

2. In the researched case, the $YBa_2Cu_3O_{7-\delta}$ superconducting thin film is under the action by the **alternate external magnetic field $H_{rf}$**, which penetrates inside the *HTS* sample on the penetration depth $\lambda$, satisfying the condition $\lambda \geq d$, where $d$ is the thickness of the $YBa_2Cu_3O_{7-\delta}$ superconducting thin film. It is necessary to note that the alternate external magnetic field $H_{rf}$ together with the alternate surface currents $J_S$ cannot generate the circular currents and magnetic vortices in the $YBa_2Cu_3O_{7-\delta}$ superconducting thin film at $H_{rf} < H_{C1}$; while the constant external magnetic field $H_e$ can generate the magnetic vortices in the $YBa_2Cu_3O_{7-\delta}$ superconducting thin film at $H_{rf} < H_{C1}$ as it was shown in the above considered case no. (*1*). The magnetic vortices begin to penetrate into the $YBa_2Cu_3O_{7-\delta}$ superconducting thin film at $H_{rf} \geq H_{C1}$, and this phenomenon is distinctive, because the magnetic vortices are concentrated in the sub-surface layer with the thickness of $\lambda$, where there is the alternate magnetic field and the alternate currents. The appearance of regions, which are analogous to the normal metal cores with the effective radius of $\xi \ll \lambda$ is enough to assume that there are the magnetic vortices in the $YBa_2Cu_3O_{7-\delta}$ superconducting thin film at the microwaves. It is possible to state that the magnetic vortices are originated, if the above condition, namely the presence of regions with the normal metal cores in the sub-surface layer in the $YBa_2Cu_3O_{7-\delta}$ superconducting thin film, is satisfied. The appearance of the normal metal cores in the sub-surface layer with the thickness of $\lambda$ in the $YBa_2Cu_3O_{7-\delta}$ superconducting thin film does not require the transposition of the magnetic fields or the currents, and shows up in the process of an annihilation of the superconducting correlations in the volume of $(\xi^2 \cdot l) \ll (\lambda^2 \cdot l)$, where $l$ is the length of the magnetic vortex. This process is not accompanied by the re-distribution of the magnetic field, but is characterized by the change of the spatial distribution of the superconducting gap $\Delta(r)$, which reduces to *0* (*zero*) at the center of the core. In this case, the magnetic field is not directly connected with the normal metal core as it is in the *Abricosov* magnetic vortex, but the magnetic field is generated by the ultra high frequency electromagnetic wave in the sub-surface layer of the high temperature superconducting thin film.

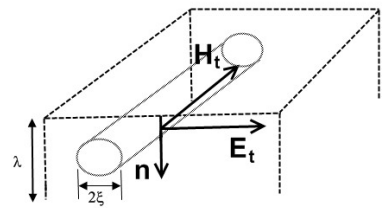

*Fig. 11*. *Normal metal core, which originates in sub-surface layer in HTS thin film along magnetic field ($\xi << \lambda$), **n** is oriented into HTS thin film.*

In this case, the normal metal cores, appearing at $H_{rf} \geq H_{C1}$, can be considered as the new independent collective excitations with the electron spectrum, which is close to the electron spectrum of the normal metal, because the energy gap $\Delta$ is suppressed in the normal metal cores. It is necessary to note that the processes of the origination and annihilation of the normal metal cores are fast enough, because the relaxation of the energy gap can have place in this region at the time period $\tau \approx \hbar/\Delta \approx 10^{-12}$ sec. This surface region in the $YBa_2Cu_3O_{7-\delta}$ superconducting thin films can be called as the vortex plasma in an analogy with the plasma, which is present in the state, when the electrons and ions are independent. Therefore, it is necessary to clarify that, in the considered case, the magnetic field is not directly connected with such physical object as the *Abricosov* magnetic vortex, but it is fully predefined by the external electromagnetic wave. Thus, at the certain amplitude of the external electromagnetic wave, the origination of the normal metal cores in the regions of an *HTS* sample thermodynamically becomes possible. These normal metal cores are oriented along the lines of magnetic field, and have the characteristic diameter of $\sim 2\xi$, and don't create the isolated magnetic vortical structures. The appearance of the normal metal cores in the $YBa_2Cu_3O_{7-\delta}$ superconducting thin films results in an additional energy dissipation by the electromagnetic wave, which takes place on the normal electron excitations. The total number of the normal metal cores in the $YBa_2Cu_3O_{7-\delta}$ superconducting thin film could increase, when the magnitude of the alternate external magnetic field $H_{rf}$ would pass the magnitude of the lower critical magnetic field $H_{C1}$. In the author's opinion, this physical mechanism leads to a nonlinear increase of the surface resistance $R_S$ at the magnitude of the microwave power $P > 8$ *dBm* as shown in Figs. 6 and 7. In the microwave resonators, the direction of the magnetic field can not only have the tangential direction in relation to the surface of a *HTS* sample, but can also have a normal component of the magnetic field, therefore the normal metal cores may have the both possible orientations. In the considered case, when $\lambda \geq d$, the full volume of the normal metal cores make a contribution to the physical mechanism of the energy dissipation in the $YBa_2Cu_3O_{7-\delta}$ superconducting thin film, and there is a thermodynamic connection between the full volume of the normal metal cores and the change of the free energy of the superconducting system or the average value of the energy gap $<\Delta>$ in the volume of a *HTS* sample at an action by the external magnetic field $H_{rf}$. Thus, let us assume that the volumetric density of the superconducting phase in the $YBa_2Cu_3O_{7-\delta}$ superconducting thin film is proportional to the $\rho$ and the volumetric density of the normal phase in the $YBa_2Cu_3O_{7-\delta}$ superconducting thin film is proportional to the $(1-\rho)$, then the surface resistance can be expressed as

$$<R_S> = R_0[\rho \cdot exp(-\Delta/kT) + (1-\rho) \cdot exp(-(\Delta-X)/kT)] =$$
$$= R_0 \cdot exp(\Delta/kT) \cdot exp((1-\rho)X/kT) = R_0 \, exp(-\Delta^*/kT), \quad (16)$$

where $X$ is the change of the energy gap in the unit of volume of $(1-\rho)$, $R_0$ is the resistance at $\Delta = 0$. The magnitude of the effective energy gap is equal to $\Delta^* = \Delta - (1-\rho)X$. Considering the connection between the energy gap and the free energy of a superconductor, it is possible to write the following expression $<R_S> = R_0 exp(-<F^*>/kT)$, where $<F^*>$ is the change of the free energy in close proximity to the lower critical magnetic field $H_{C1}$ [1]. Author can mathematically model the dependence of the surface resistance on the external magnetic field $R_S(H_{rf})$, using the measured magnitude of the critical magnetic field $H_{C1}$ in the high temperature superconductors at the microwaves [1, 20].

## Conclusion

The nonlinear surface resistance $R_S$ of the $YBa_2Cu_3O_{7-\delta}$ superconducting thin films on the *MgO* substrates was researched in the *Hakki-Coleman* dielectric resonator at the microwave signal powers from *–18 dBm* to *+30 dBm* at the ultra high frequency of *25 GHz* in the range of temperatures from *12 K* to *85 K*. The dependences of the surface resistance on the temperature $R_S(T)$ in the $YBa_2Cu_3O_{7-\delta}$ superconducting thin films at the microwaves were measured. The dependence of the surface resistance on the microwave power $R_S(P)$ in the $YBa_2Cu_3O_{7-\delta}$ superconducting thin films at the microwaves were found at the two temperatures *T = 25 K* and *T = 50 K*. The full expression for the estimation of the measurements accuracy of the surface resistance $R_S$ was derived, and the measurements accuracy was set at *1 %*. The physical mechanisms, which can be used to explain the experimental results, were discussed. It is shown that the surface resistance *Rs* can nonlinearly increase due to the transition by the sub-surface layer of the *HTS* thin film in a mixed state with the *Abricosov* and *Josephson* magnetic vortices generation at an increase of the microwave signal power *P* above the magnitude of *8 dBm*. It is assumed that some additional energy losses have place, because of the microwave power dissipation on the normal metal cores of the magnetic vortices.

[*]E-mails: dimitri.ledenyov@my.jcu.edu.au.

———————